# Hydrodynamics interactions of clusters of drops: a study of the coalescence phenomena with the finite volume method

Alejandro Acevedo-Malavé

Centro Multidisciplinario de Ciencias, Instituto Venezolano de Investigaciones Científicas (IVIC), Mérida 5101, Venezuela.

**ABSTRACT** In this work has been proposed a numerical scheme with the aim of simulate the coalescence process between water drops immersed in a continuous phase (n-heptane). This numerical scheme is based in the Finite Volume method and two different values for the initial velocity of the drops were chosen. Depending of the initial velocity of collision some scenarios emerge, such as: permanent coalescence, formation of satellite drops etc. For some snap shots the streamlines are calculated for the different process of permanent coalescence. These streamlines allow the understanding of the dynamics of the droplets immersed on the n-heptane phase. The model used for the surface tension is presented and its effects on the dynamics of coalescence of the droplets are showed, which after some time tends to the spherical form. KEY WORDS Numerical simulation, finite volume method, Navier-Stokes, coalescence, droplets.

# INTRODUCTION

Experimental studies of the binary collision of alkane droplets were carried out[6,7]. These authors found that when the Weber number is increased, the collision takes the form of a high-energy one and different types of results arises. In these references the results show that the collision of the droplets can be bouncing, grazing and generating satellite drops. A similar experimental study about binary collision of equal-size alcohol droplets has been carried-out[9] showing different stages of the collision process. An experimental study of the binary collision of water droplets for a wide range of Weber numbers and impact parameters was reported[8]. These authors identify two types of collisions leading to the drops separation, which can be reflexive separation and stretching separation. It is found that the reflexive separation occurs for head-on collisions, while stretching separation occurs for high values of the impact parameter. These authors study experimentally the border between two types of separation, and also collisions leading to coalescence. A study on the coalescence and fragmentation of mercury drops of equal and unequal sizes was published[15]. In this study these authors find out the limits for the coalescence measured in terms of the relative velocity and impact parameter. An experimental study of the effect of viscosity in the collision of drops was reported[12]. Various organic substances are used as liquid phase corresponding to a range of viscosities from 0.9 to 48 MPa s. The collision Weber numbers ranged from 10 to 420, and binary collisions of liquid drops were reported using a modified stroboscopic technique, varying the impact parameter. A study of droplet coalescence in a molecular system with a variable viscosity and a colloid-polymer mixture with an ultralow surface tension was discussed[1]. When either the viscosity is large or the surface tension is small enough, it is observed that the opening of the liquid bridge initially proceeds at a constant speed set by the capillary velocity. In the first case studied one finds that the inertial effects become dominant at a Reynolds number of about 1.5 and the neck then grows as the square root of time. In the second case one finds that decreasing the surface tension by a factor of 105 opens the way to a more complete understanding of the hydrodynamics involved. A lattice Boltzmann method-based-single-phase free surface model to study the interfacial dynamics of coalescence, droplet formation and detachment phenomena related to surface tension and wetting effects was carried out[20]. A perturbation similar to the step one in Gunstensen's color model is added to the distribution functions of the interface cells in order to incorporate the surface tension into the single-phase model. Implementations of the model are verified simulating the processes of droplet coalescence, droplet formation and detachment from ceiling and from nozzles with different shapes and different wall wetting properties. A study about the effect of glycerol on the coalescence of water drops in stagnant oil phase was discussed[19]. In this reference the authors consider the binary coalescence of water drops formed through capillaries at low inlet flow rates in an immiscible stagnant oil phase. The evolution of the coalescence process is shown in this case. An experimental investigation of binary collision of drops with emphasis on the transition between different regimes is reported[18], which may be obtained as an outcome of the collision between droplets. In this study the authors analyze the results using photographic images, which show the evolution of the dynamics exhibited for different values of the Weber number. As a result of the experiment reported here, it is proposed five different regimes governing the collision between droplets: (i) coalescence after a small deformation, (ii) bouncing, (iii) coalescence after substantial deformation, (iv) coalescence followed by separation for head-on collisions, and (v) coalescence followed by separation for off-center collisions. The study of the coalescence of two small bubbles or drops using a model for the dynamics of the thinning film is proposed[14,10]. In this report both London-van der Waals and electrostatic double layer forces are taken into account. An study about the coalescence of drops in a viscous fluid in absence of inertial effects was carried out[13]. This study is based on visual observations of drops with diameter between 20 and 100 μm that collide in a linear flow. The effect of adding a copolymer to the interface is studied. It is found that the conditions for coalescence are indicative of a complicated sequence of film configurations during the thinning process. A numerical study to simulate head-on collisions of equal-sized drops is proposed[16]. These authors solve the Navier-Stokes equations using a finite difference method. In this model the drops approach to each other until the force between them is removed before the collision occurs. When the collision occurs the fluid between the drops drains out. This leads to the formation of a thin layer that interacts with the surfaces of the drops. Then this layer is artificially removed to model the surface rupture. The problem of the film drainage between two drops and vortex formation in thin liquid films is reported[4]. These authors also develop a simple model to describe dimple dynamics[5]. The process of collision between two drops solving the Navier-Stokes equation coupled with the convection equation in order to model the interface between the drops and a gas phase was studied[17]. The simulations cover four regimes of binary collisions, which are: bouncing, coalescence, reflexive separation and stretching separation. The numerical results reported here[17] suggest that the collisions that lead to rebound between the drops are governed by the macroscopic dynamics. The authors also studied the mechanism of formation of satellite drops, confirming that the principal cause of the formation of satellite drops is the "end pinching", while the capillary wave instabilities are the dominant feature in cases where a large value of the impact parameter is employed. A Lagrangian formalism to study the coalescence phenomenon between water drops in vacuum environment is carried out[2,3]. This formalism is employed to resolve the Navier-Stokes equations by replacing the fluid with a set of particles. These particles are interpolation points from which properties of the fluid can be determined. In this study, the "Smoothed Particle Hydrodynamics (SPH)" method is applied to simulate the multiple hydrodynamics collisions and the formation of clusters of equally sized liquid drops in three-dimensional space.

## Governing equations

The governing equations can be given by the continuity Eq. (1) And the momentum Eq. (2):

$$\frac{\partial}{\partial t}(r_\alpha \rho_\alpha) + \nabla \cdot (r_\alpha \rho_\alpha V_\alpha) = \sum_{\beta=1}^{N_p} \Gamma_{\alpha\beta}, \tag{1}$$

$$\frac{\partial}{\partial t}(r_\alpha \rho_\alpha V_\alpha) + \nabla \cdot (r_\alpha(\rho_\alpha V_\alpha \otimes V_\alpha))$$
$$= -r_\alpha \nabla p_\alpha + \nabla \cdot (r_\alpha \mu_\alpha (\nabla V_\alpha + (\nabla V_\alpha)^T)) + \sum_{\beta=1}^{N_p}(\Gamma^+_{\alpha\beta} V_\beta - \Gamma^+_{\beta\alpha} V_\alpha) + M_\alpha, \tag{2}$$

where $V$ is the velocity, $p$ is the pressure, $\mu$ is the viscosity, $\rho$ is the density, $r_\alpha$ is the volume fraction of the continuous phase α, $N_p$ is the total number of phases, $M_\alpha$ describes the interphase forces acting on phase α due to the presence of other phases, $\Gamma^+_{\alpha\beta}$ represents the positive mass flow rate per unit volume from phase β to phase α. The unique term in the right hand of the equality on equation (1) occurs if the interphase mass transfer takes place.

CFX® (ANSYS® 15.0, ANSYS®, Inc. Southpointe 2600 ANSYS Drive Canonsburg, PA 15317 USA), a flow solver based on the finite volume method, was used to solve Eqs. (1) and (2). A rectangular mesh was used for this calculation with 360.000 elements. Inside the CFX module was defined the following constants: $\rho$(n-heptane)=683.8 Kg/m$^3$, $\rho$(water)=1000.0 Kg/m$^3$, $\mu$(n-heptane)=0.000408x10$^{-3}$ Kg/m.s, $\mu$(water)=1.12x10$^{-3}$ Pa.s, $r_{drop}$=15.0x10$^{-6}$ m, $\sigma$(water/n-heptane)=49.1x10$^{-3}$ N/m. Here $r_{drop}$ is the droplet radius, $\sigma$(water/n-heptane) is the interfacial tension of the system water-heptane.

## The surface tension model

The surface tension contributes a surface pressure that is the normal force per unit interfacial area "A" at points $\mathbf{x}_s$ on A. If it is considered an interface between inviscid fluids having a constant surface tension coefficient (σ), the surface force per unit interfacial area can be written as:

$$F(\mathbf{x}_s) = \sigma k(\mathbf{x}_s) \mathbf{n}(\mathbf{x}_s), \tag{3}$$

where $k(\mathbf{x}_s)$ is the curvature and $\mathbf{n}(\mathbf{x}_s)$ is the unit vector normal to A at $\mathbf{x}_s$.

If two fluids are considered, fluid 1 and fluid 2, separated by an interface at time t, the two fluids then are distinguished by some characteristic function, c(x):

$$c(\mathbf{x}) = \begin{cases} c_1, \\ c_2, \\ \langle c \rangle = (c_1 + c_2)/2, \end{cases} \quad (4)$$

where $c_1$ and $c_2$ are defined in fluid 1 and fluid 2 respectively and c at the interface. This function changes discontinuously at the interface. If one may wish to predict the motion of the interface between two incompressible fluids that can be distinguished by different densities: $\rho_1$ and $\rho_2$, then the points $\mathbf{x}_s$ on the interface A are given by:

$$\rho(\mathbf{x}_s) = \langle \rho \rangle. \quad (5)$$

If it is considered the replacing the discontinuous characteristic function by a smooth variation of fluid color c'(**x**) from $c_1$ to $c_2$ over a distance of **O**(h), where h is a length comparable to the resolution afforded by a computational mesh with spacing Δx. This replaces the boundary value problem at the interface by an approximate continuous model and reproduce the problem specification in a numerical calculation.

The interface where the fluid changes from 1 to 2 discontinuously is changed by a continuous transition. If it is considered a volume force, **F**<sub>sv</sub>(**x**) that gives the correct surface tension per unit interfacial area **F**<sub>sa</sub>(**x**<sub>s</sub>) as h→0:

$$\lim_{h \to 0} \int_{\Delta V} \mathbf{F}_{sv}(\mathbf{x}) d^3 x = \int_{\Delta A} \mathbf{F}_{sa}(\mathbf{x}_s) dA, \quad (6)$$

where the area integral is over a portion of interface area within the small volume ΔV. Additionally it is required that **F**<sub>sv</sub>(**x**) be localized so that is zero outside the interface region:

$$\mathbf{F}_{sv}(\mathbf{x}) = 0 \quad \text{for} \quad |\mathbf{n}(\mathbf{x}_s) \cdot (\mathbf{x} - \mathbf{x}_s)| \geq h. \quad (7)$$

If **F**<sub>sv</sub>(**x**) is included in the Lagrangian fluid momentum equation for an inviscid fluid in the presence of surface tension, we have

$$\rho \frac{d\mathbf{u}}{dt} = -\nabla p + \mathbf{F}_{sv}, \quad (8)$$

where ρ is the density, **u** the velocity, and p the pressure.

The smoothed function c'(**x**) varies over a thickness h through the interface by making a convolution of the characteristic function c(**x**) with an interpolation function **L**:

$$c'(\mathbf{x}) = \frac{1}{h^3} \int_V c(\mathbf{x}') \mathbf{L}(\mathbf{x} - \mathbf{x}') d^3 x'. \quad (9)$$

The function c'(**x**) is differentiable because **L** is, and

$$\nabla c'(\mathbf{x}) = \frac{1}{h^3}\int_V c(\mathbf{x}')\nabla \mathbf{L}(\mathbf{x}-\mathbf{x}')d^3x'. \tag{10}$$

The Gauss' theorem can be used here to convert the volume integral to an integral over the interface *A*:

$$\nabla c'(\mathbf{x}) = \frac{[c]}{h^3}\int_A \mathbf{n}(\mathbf{x}_s)\mathbf{L}(\mathbf{x}-\mathbf{x}_s)dA, \tag{11}$$

where [c] is the jump in color, $[c]=c_2-c_1$.

From the integral in (11), can be computed a weighted mean of the surface normal. Since **L** has bounded support, the portion of the surface contributing is $O(h^2)$. Define $\mathbf{x}_{s0}$ as the point on *A* from which the normal direction to *A*, $\mathbf{n}(\mathbf{x}_{s0})$, passes through **x**. Then $\mathbf{x}_{s0}$ is the surface point closest to **x**. The integral in (11) is, approximately,

$$\frac{1}{h^3}\int_A \mathbf{n}(\mathbf{x}_s)\mathbf{L}(\mathbf{x}-\mathbf{x}_s)dA \cong \frac{1}{h^3}\mathbf{n}(\mathbf{x}_{s0})\int_A \mathbf{L}(\mathbf{x}-\mathbf{x}_s)dA + O\left(\left(\frac{h}{R}\right)^2\right), \tag{12}$$

where R is the radius of curvature of the surface at $\mathbf{x}_{s0}$. One can bound the integral in (12) by:

$$\frac{1}{h^2}\int_A \mathbf{L}(\mathbf{x}-\mathbf{x}_s)dA \leq \mathbf{L}(\mathbf{x}-\mathbf{x}_{s0}). \tag{13}$$

As h→0, $\mathbf{L}(\mathbf{x}-\mathbf{x}_{s0})$ is zero everywhere but $\mathbf{x}=\mathbf{x}_{s0}$ and the limit of the integral of $\nabla c'(\mathbf{x})$ across the interface is given by:

$$\lim_{h\to 0}\int \mathbf{n}(\mathbf{x}_{s0})\cdot \nabla c'(\mathbf{x})dx = [c] \tag{14}$$

Then the limit h→0 of $\nabla c'(\mathbf{x})$ can be written:

$$\lim_{h\to 0}\nabla c'(\mathbf{x}) = \mathbf{n}[c]\delta[\mathbf{n}\cdot(\mathbf{x}-\mathbf{x}_s)] = \nabla c(\mathbf{x}). \tag{15}$$

This delta function can be used to rewrite $\mathbf{F}_{sa}(\mathbf{x}_s)$ as a volume integral for h=0:

$$\int_A \mathbf{F}_{sa}(\mathbf{x}_s)dA = \int_V \mathbf{F}_{sa}(\mathbf{x})\delta[\mathbf{n}(\mathbf{x}_s)\cdot(\mathbf{x}-\mathbf{x}_s)]d^3x$$

$$= \int_V \sigma k(\mathbf{x})\mathbf{n}(\mathbf{x}_s)\delta[\mathbf{n}(\mathbf{x}_s)\cdot(\mathbf{x}-\mathbf{x}_s)]d^3x. \tag{16}$$

The delta function converts the integral of **F**$_{sa}$(**x**) over a volume V containing the interface *A* to an integral over *A* of **F**$_{sa}$(**x**$_s$) evaluated at that surface. The integral relation in (16), an identity for discontinuous interfaces (h = 0), can be used to approximate interfaces having a finite thickness h when (15) is substituted for the delta function. Upon substituting (15) into (16), can be found:

$$\int_A \mathbf{F}_{sa}(\mathbf{x}_s) dA = \lim_{h \to 0} \int_V \sigma k(\mathbf{x}) \frac{\nabla c'(\mathbf{x})}{[c]} d^3 x. \tag{17}$$

If the equation (17) is compared with equation (6) the volume force **F**$_{sv}$(**x**) is identified for finite h as:

$$\mathbf{F}_{sv}(\mathbf{x}) = \sigma k(\mathbf{x}) \frac{\nabla c'(\mathbf{x})}{[c]}. \tag{18}$$

### Coalescence and fragmentation of liquid drops

In order to model the collision of liquid drops some calculations were carried out using the Volume Finite Method. It was chosen the velocity of collision with values of 0.2 m/s and 3.5 m/s which correspond to coalescence and fragmentation of water drops immersed in n-heptane continuous phase.

In Figure 1 is shown a sequence of snapshots for the collision between the water drops (blue color) immersed on a hydrocarbon continuous phase (white color). It can be seen that for a velocity of collision of 0.2 m/s the coalescence of drops is carried out, without the formation of the circular interfacial film that is reported in the literature. This behavior occurs because the liquid that drains out between the droplets has the time enough to drain and no fluid is trapped at the interface of the drops.

In fact, for this case the surface tension forces prevailing over inertial forces and the dynamics of the system water-heptane show multiple oscillations at the surface of the bigger droplet that result of the coalescence process. At t=3.44x10$^{-7}$ sec showed a little point of contact between the drops that form a bridge structure with the dispersed phase. After this period of time the evolution of the system show the increment of the radius of this bridge and a mass of the continuous phase is occluded inside the bigger drop of n-heptane. At t=2.71x10$^{-5}$ it is shown that the system recover its circular form.

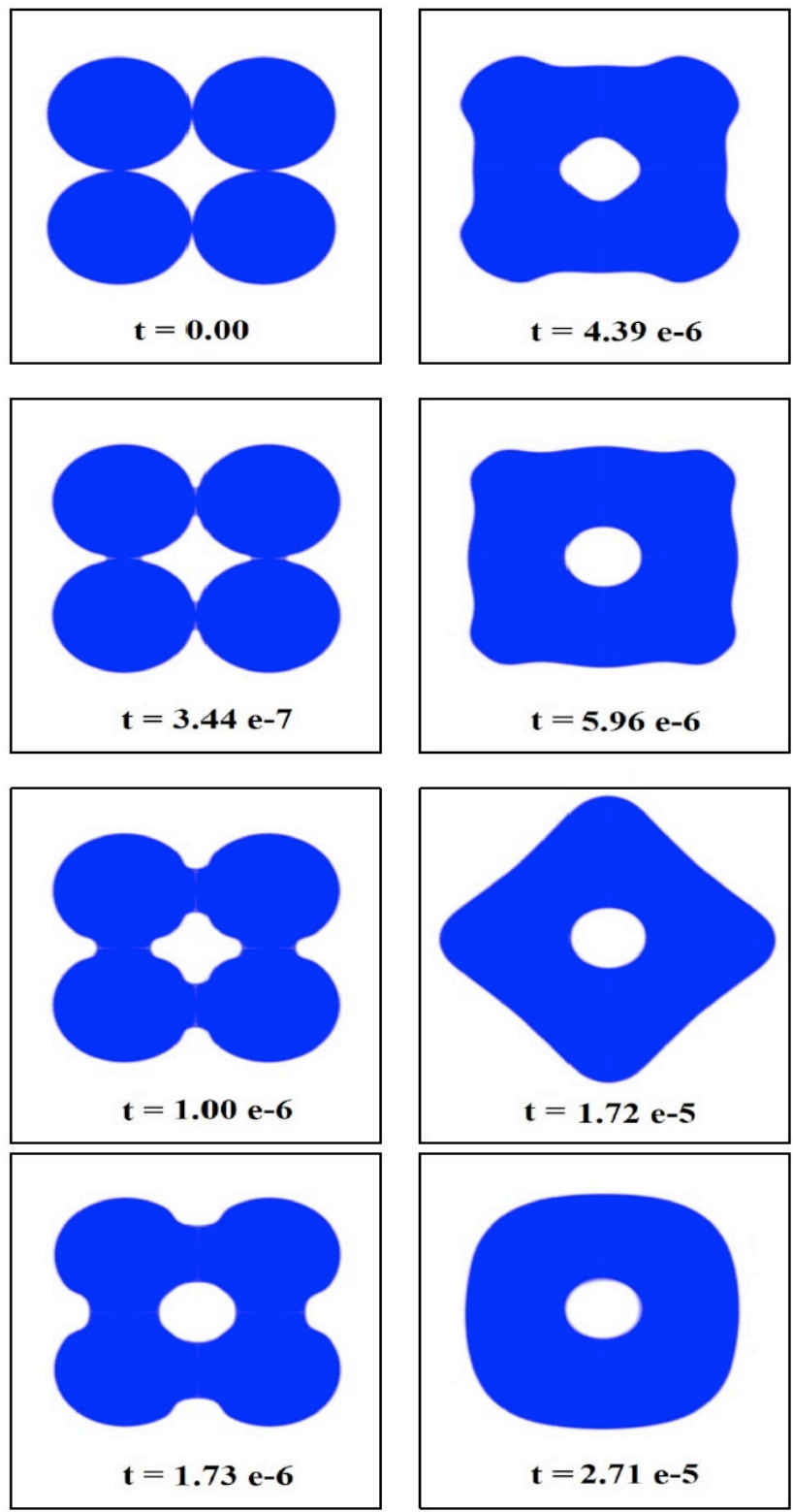

**Figure 1** Sequence of times showing the evolution of the collision between the drops with $V_{col}$ = 0.2 m/s. The time scale is given in seconds.

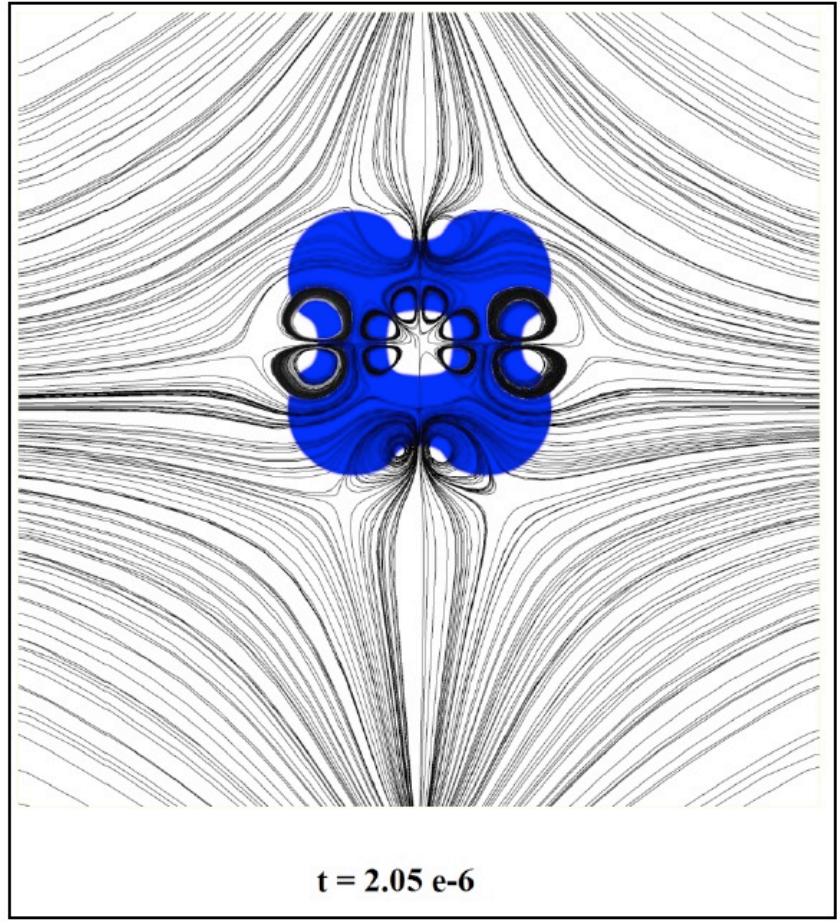

**Figure** 2 Streamlines for the system water-heptane with $V_{col}$=0.2 m/s. The time scale is given in seconds.

In Figure 2 it can be seen the internal streamlines for the system of drops at t=2.05x10$^{-6}$ sec. Once the surfaces of the drops touch between them, the internal flux of the droplets has a circular form at many regions of the system. In fact, these forms of the flux are responsible for the curvature of the bigger drop in all directions of the system. At the end of the dynamics it can be seen that the bigger drop takes a perfect circular form.

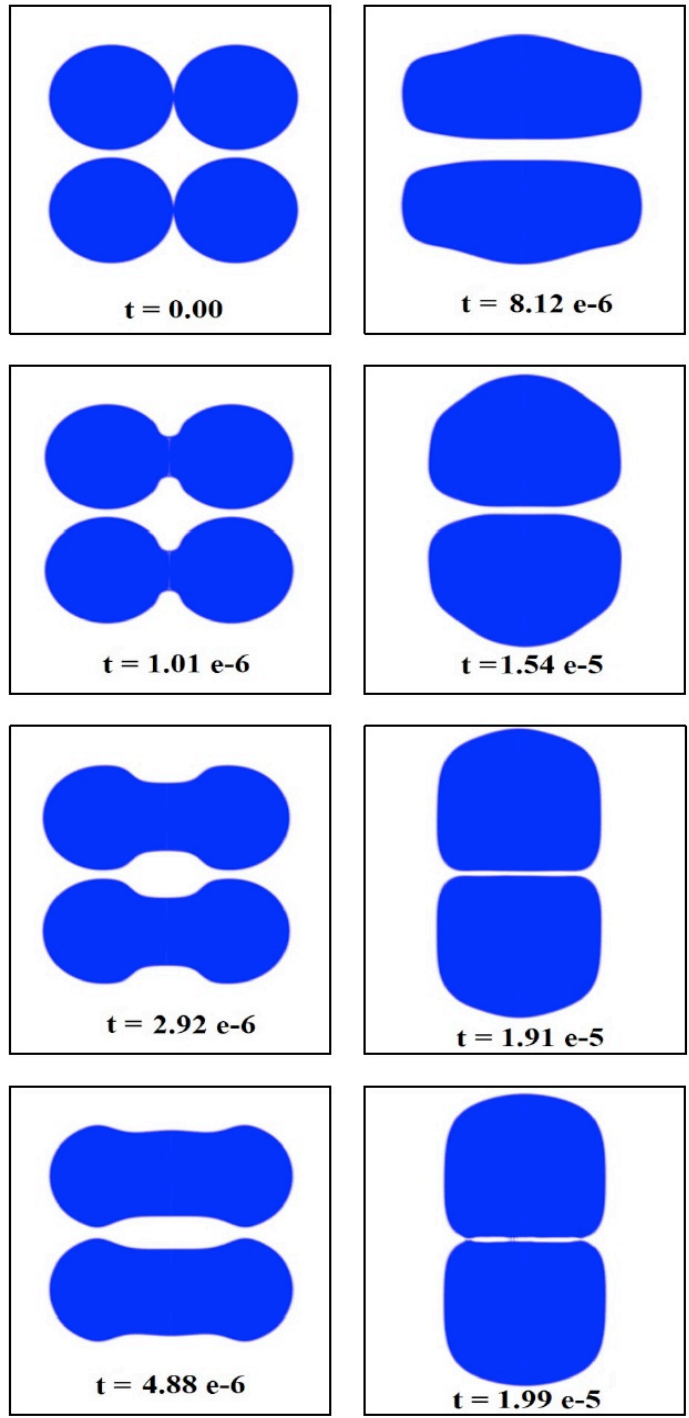

**Figure 3** Sequence of times showing the evolution of the collision between the drops with $V_{col}$ = 0.2 m/s. The time scale is given in seconds.

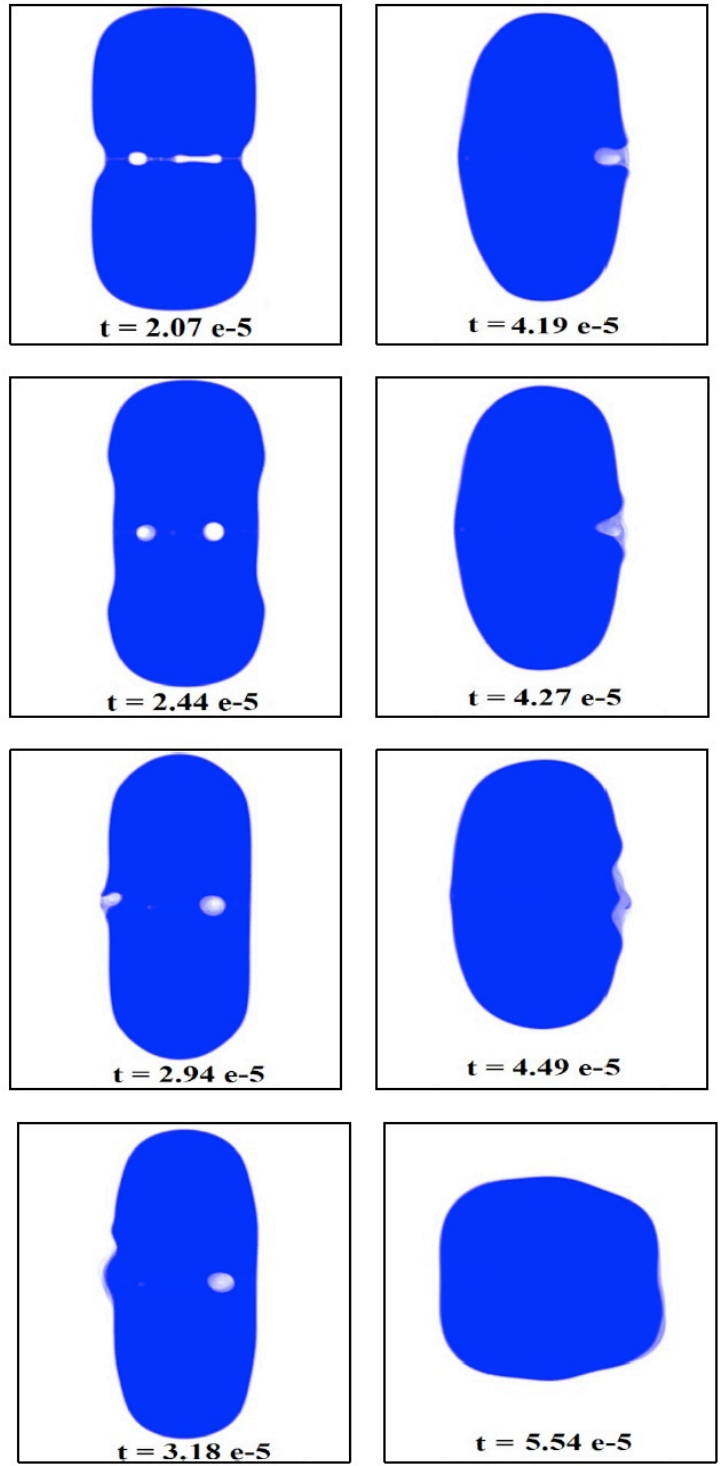

**Figure 4** Sequence of times showing the evolution of the collision between the drops with $V_{col}$ = 0.2 m/s. The time scale is given in seconds.

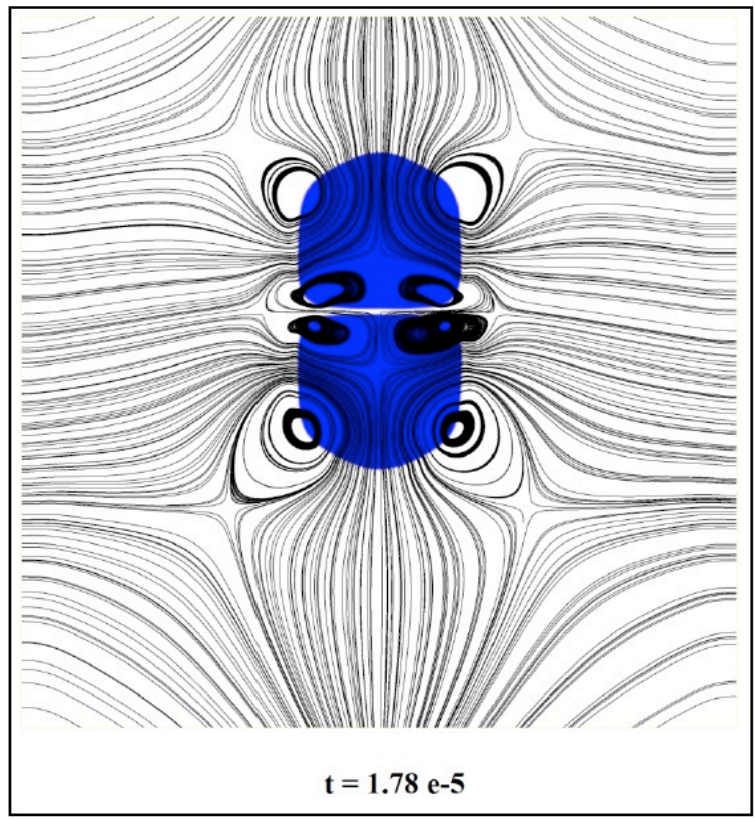

**Figure 5** Stream lines for the system water-heptane with $V_{col}$=0.2 m/s. The time scale is given in seconds.

In Figures 3 and 4 is shown the evolution of the dynamics for the collision between the water drops immersed on the hydrocarbon continuous phase. It can be seen that for a velocity of collision of 0.2 m/s the coalescence of drops is carried out with the direction parallel to "x" axis. In this stage of the dynamics there is no the formation of the circular interfacial film that is reported in the literature. Equal that first case, this behavior occurs because the liquid that drains out between the droplets has the time enough to drain and no fluid is trapped at the interface of the drops. For this value of the velocity of collision the surface tension forces prevailing over inertial forces and the dynamics of the system water-heptane show some oscillations at the surface of the bigger drop.

In Figure 5 it can be seen the streamlines for the system of drops at t=1.78x10$^{-5}$ sec. Can be observed four circular zones in the flux inside each drop and a zero flux zone at the middle of the two bigger drops. This zero flux zone is connected with the other zones of the drops where the flux is circular. Depending on the oscillations over the surface of the drops the streamlines modify its form.

In order to model the coalescence phenomenon and the formation of the interfacial film between the drops, it is chosen a collision velocity of 3.5 m/s for the water droplets immersed in the hydrocarbon phase (n-heptane). In Figures 6 and 7 is shown a sequence of times for the collision between the water drops. It can be seen that for a velocity of collision of 3.5 m/s at t=2.21x10$^{-5}$ sec the deformation of the droplet surface has begun. In the next snapshot the formation of a liquid bridge can be seen and is completely formed between the four drops at t=2.44x10$^{-5}$ sec.

At t=2.78x10$^{-5}$ sec a wave front begins to move on the surface of the drops inside the bridge. This wave front with the evolution of the dynamics made a collapse of the interfacial film between the two bigger drops, in fact at t=4.40x10$^{-5}$ sec the coalescence phenomenon has begun at the center of the film. After this, there are multiple points of coalescence and two little fractions of n-heptane are trapped in the interfacial film. This amount of hydrocarbon is expelled to the continuous phase with the evolution of the dynamics.

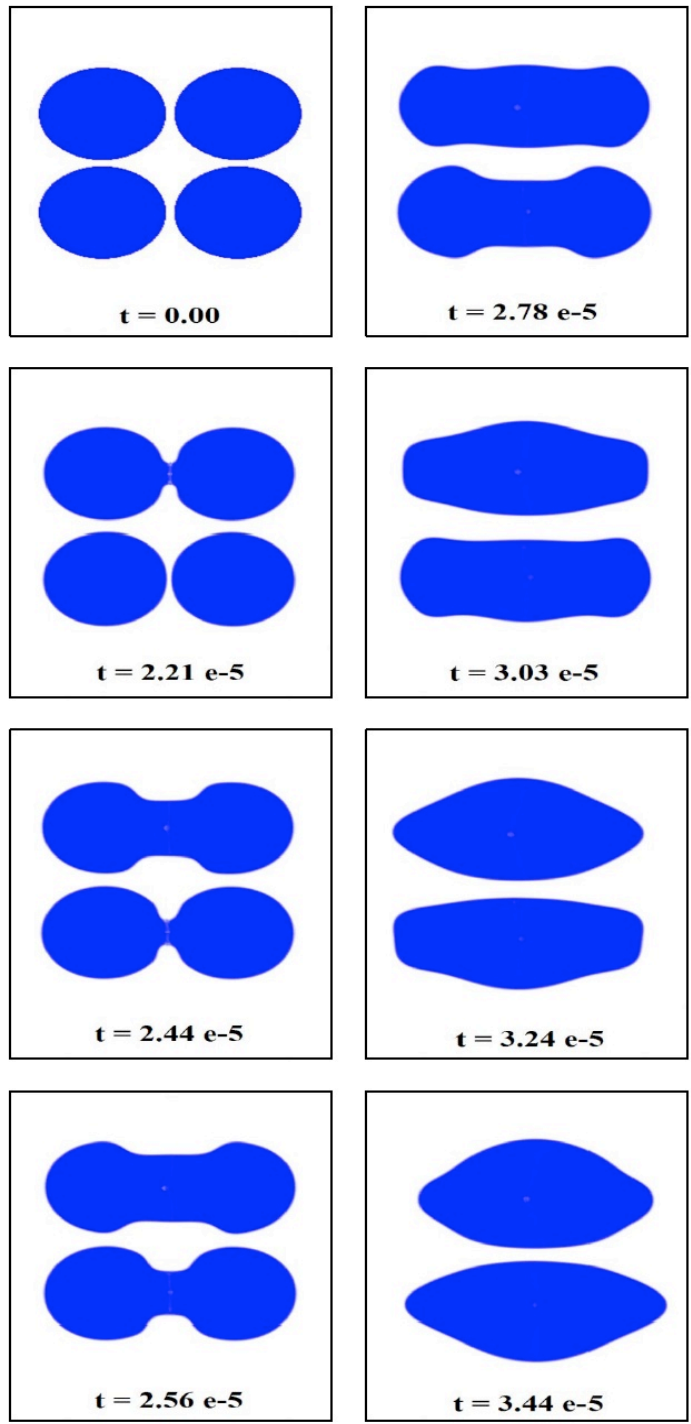

**Figure 6** Sequence of times showing the evolution of the collision between the drops with $V_{col}$ = 3.5 m/s. The time scale is given in seconds.

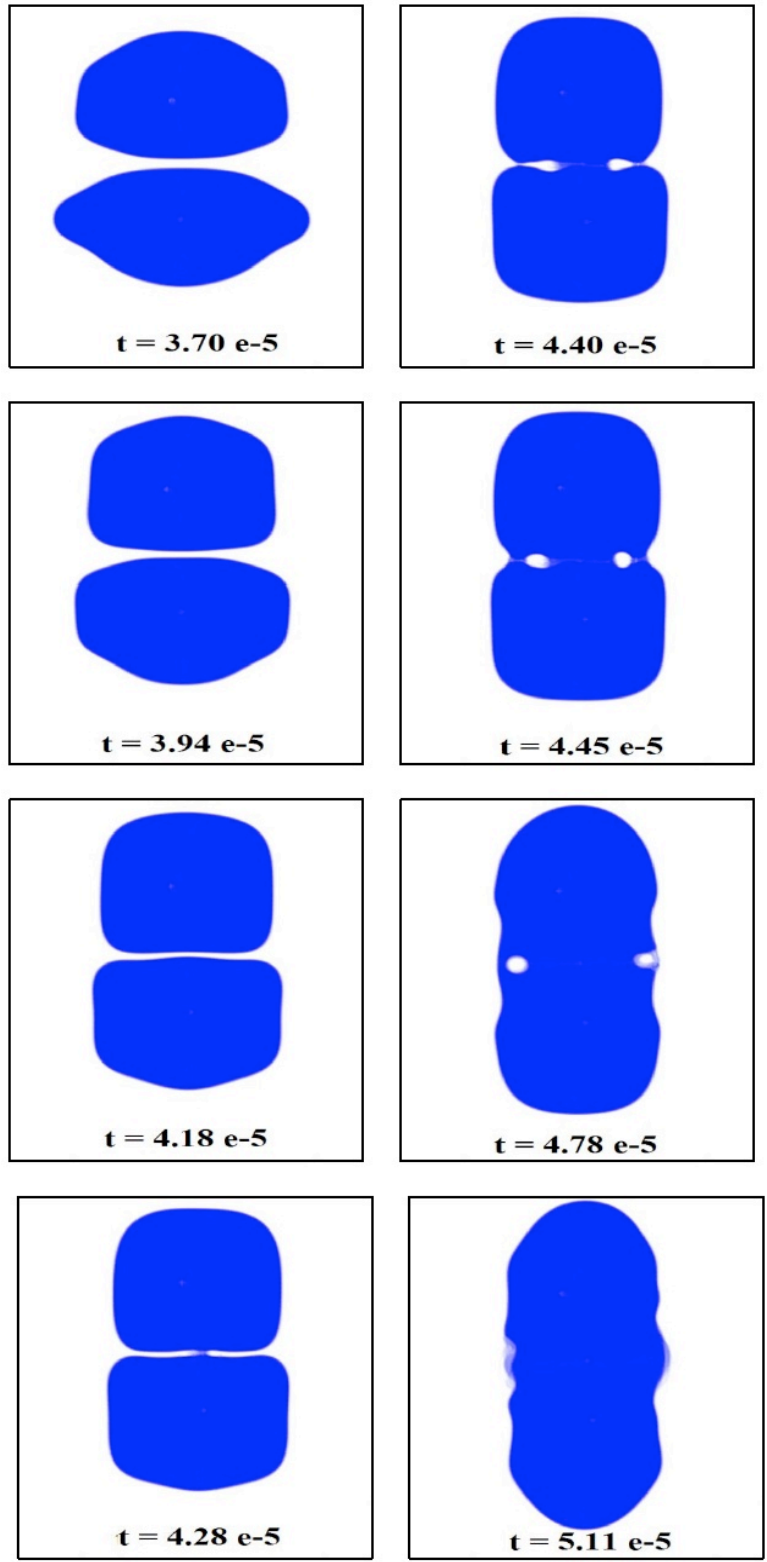

**Figure 7** Sequence of times showing the evolution of the collision between the drops with $V_{col} = 3.5$ m/s. The time scale is given in seconds.

## CONCLUSIONS

In this work a robust methodology for the study of hydrodynamical collisions of clusters of drops is presented. The equations of Navier-Stokes were resolved using the Finite Volume Method for a system composed of four water drops and the hydrocarbon (n-heptane) that represent a continuous phase. Depending on the initial conditions it was shown some different outcomes such as: permanent coalescence with the occlusion of a drop of continuous phase inside the bigger drop of water and coalescence with the formation of the interfacial film between the droplets. The streamlines are shown for the flux inside and outside of the biggest drop and it can be seen that this flux can be divided in several regions where the flux is circular when the coalescence is completed. On the other hand, when the velocity of collision was 3.5 m/s the drops stretch its surface and the formation of the interfacial film arises after this step in the dynamics of the system. The corrugations that appear in the interfacial film form some little drops of hydrocarbon inside the mass of water and these drops in the dynamics of the system coalesce with the continuous phase. The streamlines allow the understanding of why a little hydrocarbon drop is expelled out the mass of water and follow that direction.

**Correspondencia:** Dr. Alejandro Acevedo-Malavé. Centro Multidisciplinario de Ciencias, Instituto Venezolano de Investigaciones Científicas (IVIC), Mérida 5101, Venezuela. Email: alaceved@ivic.gob.ve